%% file: paper.tex
\documentclass[twocolumn]{aastex63}
\usepackage[ruled,vlined,norelsize]{algorithm2e}
\usepackage{amsmath}

\sloppypar
\shorttitle{EPRV Instrument Corrections}
\shortauthors{}

\input{customCommands}

\newcommand{\rcb}{$\rho$CrB}

\newcommand{\echellogram}{\textit{\'echellogram}}
\newcommand{\echelle}{\textit{\'echelle}}

\begin{document}
\title{Uncovering Hidden Systematics in Extreme-Precision Radial Velocity Measurements}

\correspondingauthor{Lily L.\ Zhao}
\email{lilylingzhao@uchicago.edu}

\author[0000-0002-3852-3590]{Lily L.\ Zhao}
\thanks{NASA Sagan Fellow} 
\affil{Department of Astronomy \& Astrophysics, University of Chicago, Chicago, IL, USA}

\author[0000-0003-2221-0861]{Debra A. Fischer} 
\affiliation{Department of Astronomy, Yale University, 219 Prospect St., New Haven, CT 06511, USA}

\author[0000-0002-4974-687X]{Andrew E. Szymkowiak} 
\affiliation{Department of Physics, Yale University, 217 Prospect St., New Haven, CT 06511, USA} 
 
\author[0000-0002-9873-1471]{John M. Brewer} 
\affiliation{San Francisco State University University, 1600 Holloway Ave., San Francisco, CA 94132, USA} 

\author[0000-0003-4450-0368]{Joe Llama} 
\affiliation{Lowell Observatory, 1400 W. Mars Hill Rd., Flagstaff, AZ 86001, USA}

\begin{abstract}
We identify and correct for small but coherent instrumental drifts in seven years of radial velocity data from the EXtreme PREcision Spectrograph (\expres).  The systematics are most notable for the six months before and after 2022~January, when \expres\ experienced larger temperature variations, and we see a systematic trough-to-peak amplitude of 2.8 \ms\ in the radial velocities.  This is large enough to mimic or obscure planetary signatures. To isolate and correct these effects, we develop a suite of diagnostics that track two-dimensional \echellogram\ shifts, scalings, and rotation, as well as line bisector spans (LBS) derived from laser frequency comb (LFC) lines. By combining these empirical tracers with instrument telemetry in a multi-dimensional regression, we reduce the \expres\ instrument trend traced with solar RVs from an RMS of 1.32~\ms\ to 0.43~\ms, a 67\% improvement, and the aggregate of twelve chromospherically quiet stars show a 26\% reduction in velocity scatter. Our injection–recovery simulations further demonstrate a doubling in sensitivity to low-amplitude planetary signals after correction. When applied to the stellar time series of $\rho$~Coronae~Borealis (\rcb), the correction removes a spurious planet~d signal, restoring the integrity of the data.  These results highlight the need for long-term monitoring and multi-dimensional calibration diagnostics on the path toward true centimeter-per-second precision in next-generation EPRV instruments.
\end{abstract}

\keywords{techniques: spectroscopic -- techniques: radial velocities -- methods: data analysis -- methods: statistical}

\section{Introduction}\label{sec:introduction}

Planet-hunting spectrographs are now enabling sub-meter-per-second radial velocity (RV) measurements \citep{pepe2004, cosentino2012, pepe2013, schwab2016, jurgenson2016, seifahrt2018, gibson2018}.  This level of precision allows for the detection of smaller-amplitude signals and lower-mass planets \citep[e.g.][]{faria2022, gonzlezhernandez2024, basant2025}.  The latest generation of instruments are all variants of ultra-stabilized \'echelle spectrographs that yield very high-fidelity spectra.

\textit{\'Echelle} spectrographs disperse light in two perpendicular directions, forming a two-dimensional projection of the spectrum on the detector known as an \echellogram. The resulting horizontal ``bands,'' appear like the rungs of a ladder; indeed, \textit{\'echelle} is French for ``ladder.''  The horizontal bands are referred to as \echelle\ orders.  The bands are distinct from one another, and so the \echelle\ order number and pixel position across the order is often used to uniquely define a position within the \echellogram.

The precision achieved by a spectrograph depends not only on its hardware but also on the extraction pipeline, which converts the two-dimensional \echellogram\ into a one-dimensional spectrum. This process involves tracing each order across the detector and extracting its corresponding flux profile.

A central step in this process is wavelength calibration, the mapping between detector pixels and physical wavelengths. For radial-velocity (RV) planet searches, where precision at the 10~\cms\ level is required, the fidelity of this calibration is paramount. Emission-line calibration sources give rise to reference wavelengths that can be fitted by low-order polynomials.  More recently, non-parametric models have been enabled by laser frequency combs and other dense calibrators~\citep[e.g.,][]{probst2014, zhao2021}.

Current approaches implicitly assume that calibration sources perfectly trace all instrumental variations and that these variations affect emission and absorption lines in the same way. In reality, any wavelength-dependent deviations from these assumptions remain uncorrected and can propagate directly as scatter in stellar RV measurements. Moreover, asymmetric instrument profiles (IPs) can bias the measured positions of calibration lines.  For example, recent work with \espresso\ calibration data showed that explicitly modeling IP asymmetries led to a fivefold improvement in wavelength calibration \citep{schmidt2024}.

Such uncorrected effects contribute to \textit{instrument drift}, a shift in the spectrograph's RV zero point relative to a reference calibration. Historically represented as a single scalar offset, instrumental drift is now understood to vary across the detector, motivating two-dimensional drift maps~\citep{dumusque2021}. These maps reveal spatial structure in residual calibration errors that can degrade long-term RV stability.  

In this work, we identify a coherent instrumental trend in long-term \expres\ observations that is not corrected by the standard pipeline calibration. The trend became apparent only after several years of data accumulation and careful temporal binning. In \S\ref{sec:data} we introduce the \expres\ data set; in \S\ref{sec:diagnostic} we describe binning choices and diagnostic tracers of systematics; in \S\ref{sec:methods} we develop a simple correction model; and in \S\ref{sec:results} we present the corrected results. We conclude with discussion and implications in \S\ref{sec:discussion} and summarize in \S\ref{sec:conclusion}.

\section{Data}\label{sec:data}

We make use of data taken by \expres, a stabilized, fiber-fed \'echelle spectrograph~\citep{jurgenson2016} commissioned on the 4.3~m Lowell Discovery Telescope~\citep[\ldt;][]{levine2012} near Flagstaff, Arizona, in 2019~January. As part of the 100~Earths Survey~\citep{brewer2020}, \expres\ has acquired observations spanning seven years at the time of this writing. When considering calibration sources alone, \expres\ exhibits an instrumental stability of 4--7~\cms~\citep{blackman2020}, while on-sky observations achieve a demonstrated precision of 25-30~\cms\ for SNR$>$200 \citep{petersburg2018}.  The spectrograph delivers a median resolving power of $R \approx 137{,}000$, aided by a rectangular optical fiber that approximates the behavior of a slit.  During the initial alignment of \expres, care was taken to align the projection of this pseudo-slit with the columns of the detector.

For wavelength calibration, \expres\ utilizes a Menlo Systems laser frequency comb (LFC)~\citep{wilken2012, molaro2013, probst2014, probst2020, milakovich2020}, which produces evenly spaced lines across the $\sim$490--730~nm range. LFC exposures are obtained every 20--45~minutes during both nighttime and daytime operations, with three consecutive LFC frames included in the beginning- and end-of-night calibration sequences. Because of \expres's rectangular fiber, individual LFC lines are modeled using a super-Gaussian function,
\begin{equation}
I(x) = A \exp{\!\left[-\left(\frac{(x-\mu)^2}{2\sigma^2}\right)^{p}\right]},
\end{equation}
where $\mu$ represents the measured line center.  

Environmental monitoring is provided by an array of commercial platinum resistive temperature detectors (RTDs). A Lakeshore sensor records the temperature of the Invar optical bench within the vacuum enclosure, while an Omega sensor measures the exterior temperature of the vacuum vessel. A third Omega sensor is mounted on the concrete slab that supports \expres, which is seismically isolated from the surrounding telescope structure. These will hereafter be referred to as the optical-bench, vacuum-enclosure, and slab temperatures, respectively. Each sensor records measurements approximately every 2~minutes and 24~seconds with milli-kelvin precision.  Hardware issues temporarily halted the reading of temperature measurements for the beginning of 2025.

\subsection{Solar Data}
\expres\ has also been collecting disk-integrated solar observations (i.e., Sun-as-a-star observations) during the daytime since 2020~November. With solar data, it is possible to remove the known Doppler shifts from orbiting Solar System planets, leaving only instrumental systematics and RV variations due to solar surface variations. Light from a dedicated solar telescope is fed into the front-end module of \expres, after which the light proceeds through the same optical path as nighttime observations~\citep{llama2024}.

The diagnostic power of solar observations has motivated the addition of solar feeds to several EPRV spectrographs, enabling more robust characterization of instrumental systematics through cross-comparison of solar data among facilities. Here, we also make use of optical-to-near-infrared (380--1046~nm) solar data from the \neid\ instrument~\citep{schwab2016, halverson2016, lin2022}. Similar to \expres, \neid\ is a stabilized, fiber-fed spectrograph located on the \wiyn\ 3.5~m telescope\footnote{The \wiyn\ Observatory is a joint facility of the NSF’s National Optical-Infrared Astronomy Research Laboratory, Indiana University, the University of Wisconsin–Madison, Pennsylvania State University, and Purdue University.} at Kitt Peak National Observatory near Tucson, Arizona. \neid\ began solar observations in 2020~December. Because both instruments are located in Arizona, there is substantial overlap between the \expres\ and \neid\ solar data sets.

With solar observations from two independent instruments, we can difference out stellar signals, effectively isolating instrumental systematics.  To compare \expres\ and \neid\ observations, we bin the Solar data from each instrument onto the same time stamps.  Points are binned in 16.2~minute windows (approximately three times the solar $\nu_{\mathrm{max}}$) using a weighted average, where weights depend on both temporal distance from the bin center and measurement uncertainties~\citep{zhao2023-10}.  This results in 11,395 binned timestamps with values for both \expres\ and \neid\ ranging from 2021~January to 2025~June, with gaps when either \expres\ or \neid\ were not observing.

The 2025 \neid\ and \expres\ data are systematically offset from each other, with all \neid\ 2025 data exhibiting greater RVs relative to \expres.  Here, we are only interested in isolating instrumental systematics, so we simply subtract a single constant offset from the 2025 \neid\ binned data.  This offset was determined using only the 2025 time stamps for which there are both \neid\ and \expres\ binned values.  The 2025 \neid\ data was shifted by 3.86~\ms\ so that the median of the binned 2025 \neid\ data matches the median of the binned 2025 \expres\ data.  We then directly difference the resultant binned RVs to get residual RVs, which we will denote $\Delta$ solar RVs.

Because the two instruments are observing the same Sun, differencing the binned data produces a time series with largely instrument systematics.  Of course, these residuals will contain instrument systematics from both \expres\ and \neid.  Some solar signals may also persist due to differences in the two instruments, for example the longer wavelength range of \neid\ and differences in each instrument's pipelines.

\begin{figure*}[t!]
\centering
\includegraphics[width=.7\textwidth]{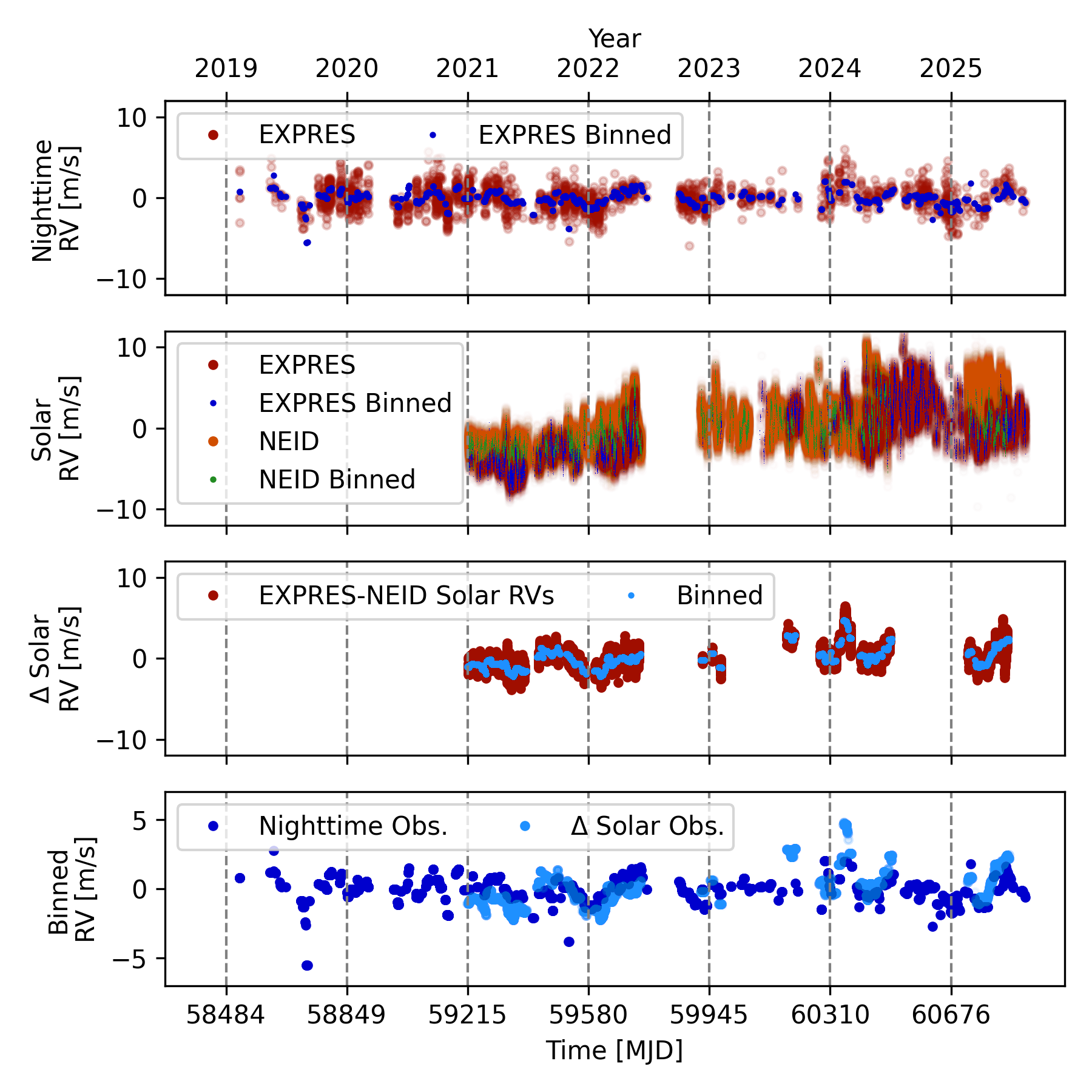}
\caption{\expres\ instrument trend in nighttime and $\Delta$ solar data. 
\textbf{Row One}: \expres\ RVs from twelve low-RMS stars (red); dark-blue points show these RVs median-filtered over 11~days. 
\textbf{Row Two}: \expres\ (red) and \neid\ (orange) solar RVs, together with each time series binned to shared time stamps (dark blue and green, respectively). 
\textbf{Row Three}: Residuals between the \expres\ and \neid\ solar data ($\Delta$ solar in red) and the same residuals median-filtered over 11~days (light blue). 
\textbf{Row Four}: Zoom in on the 11-day median-filtered points from nighttime (dark blue) and $\Delta$ solar (light blue) data. 
This corresponds to the same data shown in dark blue in row one and light blue in row three. 
Vertical dashed gray lines mark the beginning of each calendar year from 2019 to 2025.}
\label{fig:trend}
\end{figure*}

\subsection{Instrumental Trend}\label{ssec:inst_trend}

Figure~\ref{fig:trend} (top row) shows RVs for a subset of stellar targets monitored by \expres\ (red), along with the same points median-binned (dark blue).  The subset is composed of twelve stars from the \expres\ 100~Earths Survey with spectral types ranging from F9 to K3.  They were chosen on the basis of exhibiting an RMS scatter below 2.5~\ms, thereby excluding targets with significant planetary or stellar variability.  The RVs from each target are measured using a template-based, chunk-by-chunk code \citep{petersburg2020} and median-centered.  Points are binned using a median filter with an 11-day width, emphasizing signals common to all targets (as discussed further in \S\ref{ssec:binning}).

Rows two and three of Figure~\ref{fig:trend} show the $\Delta$ solar data. Row two presents the raw \expres\ (red) and \neid\ (orange) solar RVs together with the data binned to shared time stamps in blue and green respectively.  Row three of Figure~\ref{fig:trend} displays the difference between the \expres\ and \neid\ solar RVs (red), which largely removes shared solar signals. The light-blue points show these residuals median-binned over 11~days, analogous to the nighttime stellar data in the top row.

The bottom row of Figure~\ref{fig:trend} highlights only the binned \expres\ stellar data and the binned $\Delta$ solar RVs. These binning choices were optimized to most clearly reveal the \expres\ instrumental trend. The dark-blue points are identical to those in the top row, while the light-blue points correspond to the residuals from row three. Note, to highlight the trend, the y-axis range for this bottom row is smaller than the preceding subplots.  The trends in the nighttime and solar data closely track one another, with both showing a distinct dip reaching a minimum around MJD~59580, corresponding to early~2022.


\section{Identifying and Tracing Instrument Systematics}\label{sec:diagnostic}
The presence of instrumental systematics can be identified by isolating trends that appear across all observations, regardless of target. A simple approach is to bin observations over multiple targets, which averages out astrophysical variability while preserving any instrument-wide behavior. Below, we describe how we optimized this binning framework to reveal a global trend in \expres\ data and provide guidance on averaging in the presence of genuine Doppler shifts from orbiting planets.

It is also instructive to measure potential sources of instrumental systematics not addressed by standard pipeline corrections. These include asymmetries in the instrument profile (IP) and wavelength-dependent shifts at the level of the \echellogram. We present a measurement of IP asymmetry using laser frequency comb (LFC) lines, as well as a suite of metrics that trace positional changes in the \echellogram. These diagnostics can be used both to identify and to model out instrumental systematics.  

\subsection{RV Binning}\label{ssec:binning}
Instrumental systematics underlie all RV measurements from a spectrograph, regardless of the target being observed. Detecting such systematics can be challenging because observations also capture astrophysical variability, including Doppler shifts from orbiting planets and stellar surface variation. The simplest strategy for isolating instrumental trends is to average over non-instrumental variability by binning RV measurements across multiple targets.

Although the process of binning is conceptually simple, care must be taken to optimize the binning parameters, especially at the precision level required for extreme-precision radial velocity (EPRV) studies. Bins that span too large a time range will smooth over instrumental variations, while bins that are too narrow will fail to average out astrophysical or photon-noise contributions.

\begin{figure}[h!]
\centering
\vspace{-14pt}
\includegraphics[width=.35\textwidth]{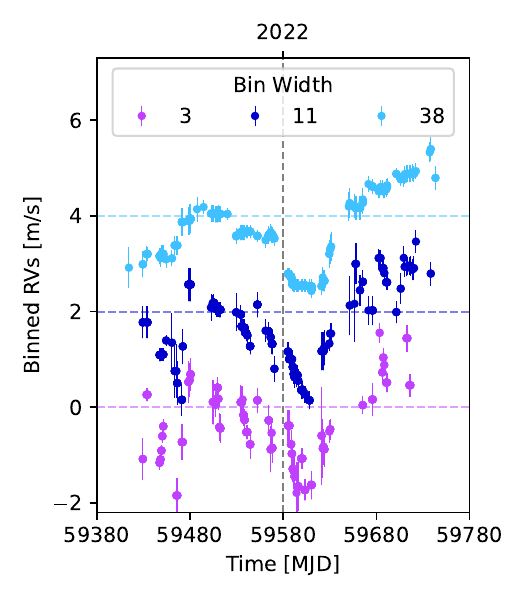}
\vspace{-8pt}
\caption{{\expres\ nighttime data from 2021~June to 2022~July median-filtered with widths of 3~days (purple), 11~days (blue), and 38~days (light blue). Each curve is vertically offset, with its RV zero point shown as a horizontal dashed line of the corresponding color. The vertical dashed gray line marks the beginning of~2022.}}
\vspace{-16pt}
\label{fig:binwid}
\end{figure}

\begin{figure*}[t!]
\centering
\includegraphics[width=\textwidth]{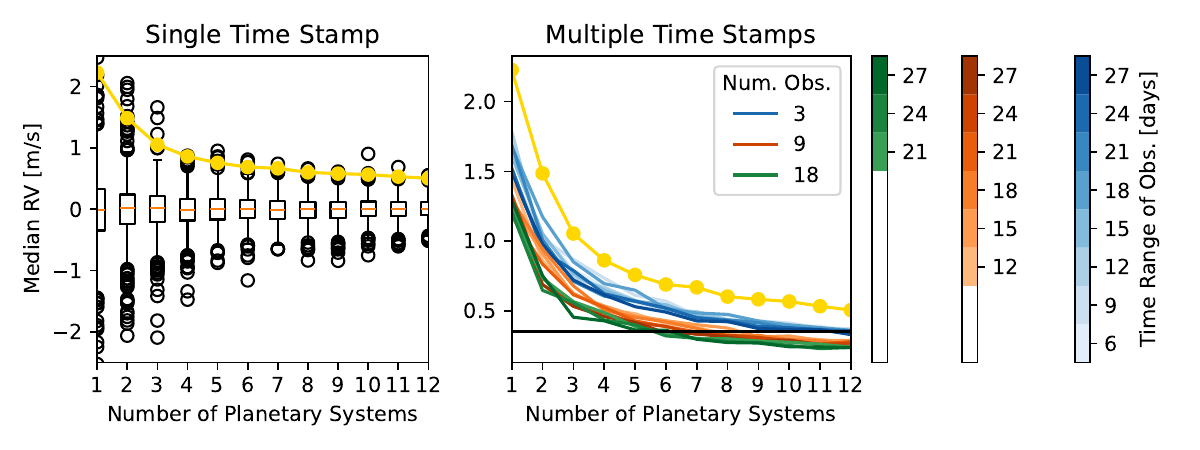}
\caption{Scatter from binning over RVs of simulated planetary systems. 
\textbf{Left, Single Time Stamp}: Distribution of the median RV as a function of the number of planetary systems sampled at a single epoch. The yellow curve traces 3$\sigma$ for each distribution. 
\textbf{Right, Multiple Time Stamps}: 3$\sigma$ of the median RV for different numbers of planetary systems sampled over multiple nights (different colors) and time ranges (color saturation). The yellow curve reproduces 3$\sigma$ for the single-epoch simulations shown in the left panel.}
\label{fig:binning}
\end{figure*}

Figure~\ref{fig:binwid} shows nighttime \expres\ data binned with various window widths. We focus on the period from 2021~June to 2022~July, during which the \expres\ data exhibit a distinct dip (see Figure~\ref{fig:trend}).  We use a time-based median filter wherein each point is assigned the median of all points within a given time range.  The standard error of this median is used as the error for each binned point.  The shortest bin width shown in Fig.~\ref{fig:binning} of 3~days (purple) produces an overly noisy time series; consecutive points show significant variation beyond what is expected given the error bars of each binned point.  The longest width of 38~days (light blue), on the other hand, blurs out features such as the base of the dip that is seen in the other binned curves.  Many consecutive observations also have close to the exact same value.

By trying a range of bin widths (not all shown here) between the two extremes of too noisy and too smooth, we found that an 11-day window (dark blue) clearly captures instrumental trends of interest.  It is difficult to determine one true ``optimal'' bin width when the exact nature of the instrumental signal is unknown.  Exploring a range of bin widths can reveal different signals that may otherwise be obscured through noise or smoothed away at different bin widths.

To evaluate how planetary signals may be affected by binning, we considered simulated RV surveys. Planetary systems were drawn from catalogs generated by \project{SysSimExClusters}, which synthesizes multi-planet architectures based on \textit{Kepler} occurrence rates and intra-system correlations~\citep{he2019, he2020, he2021}. We then computed the expected RVs for these systems, assigning random times of periastron and adding white noise at the level of 35~\cms.

As a baseline test, we simulated expected RVs from one to twelve randomly selected systems at a single epoch. Taking the median RV across all systems provides an analogue to a bin that combines observations from that number of targets. The left panel of Figure~\ref{fig:binning} shows box-and-whisker plots of the distribution of combined RVs as a function of the number of planetary systems. As is to be expected from the central limit theorem, the median of each distribution remains close to zero.

More notable is the steady decrease in scatter as the number of planetary systems increases. The yellow curve in Figure~\ref{fig:binning} traces the 3$\sigma$ width of each distribution, which follows an approximate power-law decline and plateaus near six systems.

We next examined how the combined RV scatter depends on (1) the total time span of the observations and (2) the number of observations within that span (where we equate each night with one observation). For an RV survey, the time span would correspond to the bin width, while the number of observations/nights corresponds to cadence.  These simulations capture how bins that sample different fractions or phases of planetary orbits can produce different averaged values.  For each simulation, the time stamps of the RVs from each system were randomly selected within the chosen range and then scattered randomly within $\pm$3~hours of midnight.  We generate one observation per night.  Independent randomized observation times were generated for each system, and the final binned value was set as the median of all RVs across all systems and epochs.

The right panel of Figure~\ref{fig:binning} presents these results. Curves of different colors represent different cadences, and color saturation indicates the total time span. For instance, the lightest blue curve represents three observations distributed across six nights for each target, while the darkest green represents 18~observations across 27~nights for each target. The yellow curve provides a reference to the single-epoch case shown in the left panel. As expected, higher cadence and longer bin widths both reduce scatter, although the improvement is modest. Variations in total time span lower the scatter by only a few~\cms, whereas increased cadence yields a larger reduction.

These simulations do not include stellar variability, which is unlikely to add coherently across targets. In contrast to planetary signals, which are phase-coherent, stellar activity produces stochastic or quasi-periodic variations that naturally average out over multiple stars and time baselines.

Even with only two planetary systems, 3$\sigma$ is less than 1~\ms. Within the \expres\ survey, only 2.75\% of binned points are based on observations from a single target. The coherent trend observed in the binned \expres\ data has an amplitude greater than 1~\ms, suggesting that the trend is unlikely to arise from true planetary signals alone.

\subsection{LFC Line Bisector (LBS)}\label{ssec:lbs}

Using laser frequency comb (LFC) lines to measure a line bisector span (LBS) provides a means to trace temporal changes in the asymmetry of the instrumental profile (IP). This approach is feasible because the individual LFC modes are much narrower than the resolving power of \expres\ and can therefore be approximated as delta functions that are convolved with the spectrograph’s IP. The resulting LBS measurement is conceptually analogous to the bisector inverse span~\citep[BIS;][]{queloz2001}, which is commonly used to quantify asymmetry in the cross-correlation function (CCF) between a stellar spectrum and a absorption line mask.

\begin{figure}[h!]
\centering
\includegraphics[width=.45\textwidth]{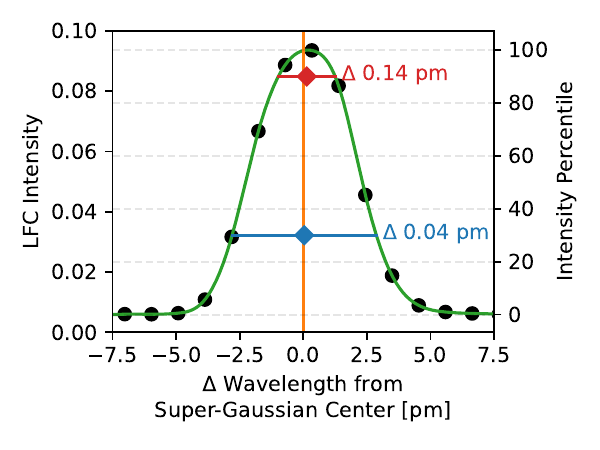}
\caption{Demonstration of the LFC line bisector span (LBS) measurement for a single LFC line. The observed LFC spectrum is shown as black points; the oversampled instrument profile (IP) is shown in green. The line is centered by the mean of a super-Gaussian fit (vertical orange line). The bisector positions at the 30th (blue) and 90th (red) percentile flux levels are indicated by diamonds. The LBS for this example line is 0.1(=0.04-0.14).}
\label{fig:lbs}
\end{figure}

For the bisector analysis, we use a set of 20 consecutive comb modes spanning 5838--5841~\AA, near the center of the full LFC wavelength range.  Given that each LFC line has a different pixel sampling, we use a cubic spline to interpolate each line and generate a smooth representation of the IP. A local baseline, determined from a linear fit across the minimum of the troughs between each of the 20 consecutive combe modes, is subtracted from each LFC line to account for the varying LFC background.  Each line is then normalized by the line’s peak flux.  We only consider LFCs with intensity above a given threshold in the wavelength range being analyzed.  For these lines, we identify the 30th and 90th percentile flux levels and determine the corresponding wavelengths at the left- and right-hand intersections of the IP for each percentile. The midpoint between the left and right wavelengths defines the line bisector at that flux level. The bisector span is quantified as the difference between the bisector midpoints at the low and high flux levels, thereby providing a direct measure of IP asymmetry. A schematic illustration of this calculation is shown in Figure~\ref{fig:lbs}.

A bisector span is computed independently for each of the 20 LFC lines between 5838--5841~\AA; the average of these values defines the representative LBS for the spectral region. The standard deviation among the 20 individual measurements is adopted as the uncertainty in the LBS.

\begin{figure*}[t!]
\centering
\includegraphics[width=.68\textwidth]{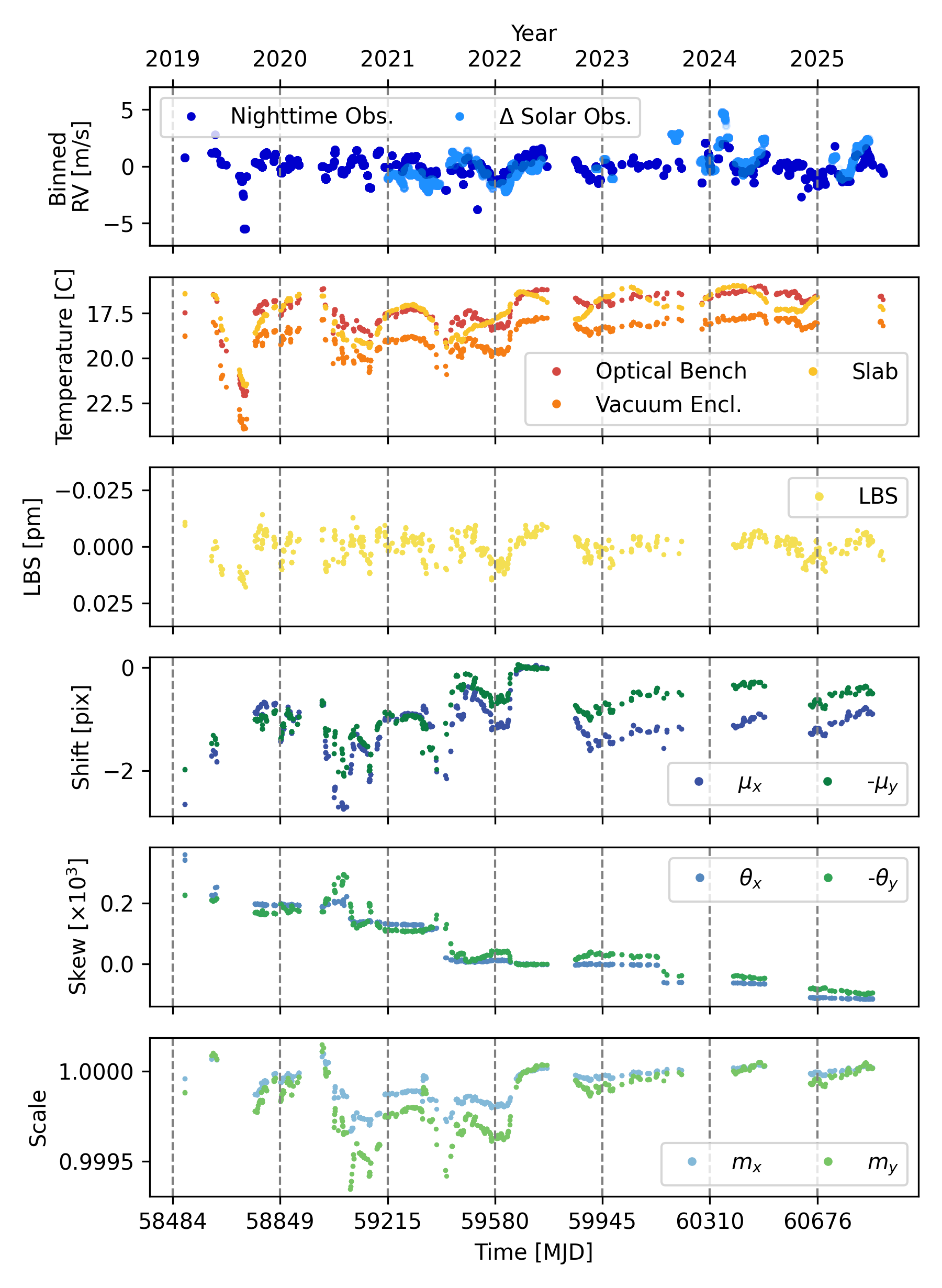}
\caption{Time series of various tracers of the instrument state. From top to bottom: the measured instrument trend (for reference), temperature measurements, the LFC line bisector span (LBS) tracing IP asymmetry, and \echellogram\ shifts, skews, and scalings in both the dispersion ($x$) and cross-dispersion ($y$) directions. Vertical dashed gray lines mark the beginning of each calendar year from 2019 to 2025.}
\label{fig:housekeeping}
\end{figure*}

The resulting LBS time series (Figure~\ref{fig:housekeeping}, row three) indicates that the instrument profile exhibits small but measurable asymmetry that varies over time. This asymmetry is expected to be imprinted on the stellar absorption lines in science observations, introducing a time-dependent effect not captured when fitting LFC line positions using symmetric profiles.

\begin{figure*}[t]
\centering
\includegraphics[width=.8\textwidth]{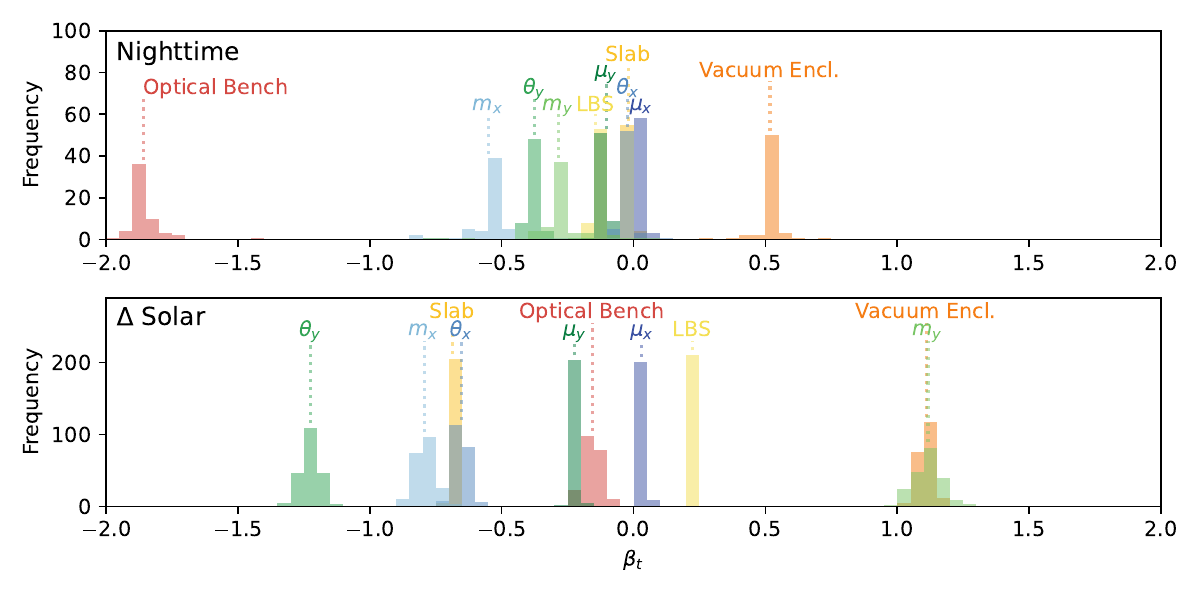}
\caption{Distribution of best-fit $\beta_t$ coefficients across all training and validation sets for nighttime data (top) and $\Delta$ solar data (bottom). Each colored histogram corresponds to a different instrumental tracer $x_t$.}
\label{fig:beta}
\end{figure*}

\subsection{Echellogram Position}\label{ssec:echellogram}

To measure changes in the \echellogram\ position, we determine the center coordinates of each LFC line in 2D space on the detector by fitting the reduced (but not yet extracted) data with a rectangle convolved with a Gaussian.  Lines are fit from three different \textit{\'echelle} orders—120, 104, and 86—which lie near the bottom, middle, and top of the LFC range (orders 122–85). These correspond approximately to wavelength ranges of 5111–5131~\AA, 5906–5921~\AA, and 7157–7161~\AA, respectively.  Incorporating lines that span the full detector helps trace the expected subtle changes across the full \echellogram.  If too many LFC lines return a poor fit (e.g.\ due to low SNR), then we do not attempt the following calculations.  By operating directly on the reduced two-dimensional data, prior to spectral extraction, we retain sensitivity to changes in both the $x$ (dispersion) and $y$ (cross-dispersion) directions. Analyses performed on extracted one-dimensional calibration spectra inherently neglect the $y$ dimension.

We measure shifts ($\mu_{x/y}$), scalings ($m_{x/y}$), and skews ($\theta_{x/y}$) of the \echellogram\ relative to a template LFC.  Here, the template LFC was chosen to be a relatively high SNR LFC close to the center of the time range.  Becasue for all values it is only the relative difference that matters, the exact choice of template is not critical.  The mean $x$ and $y$ positions across all lines is used to find the $\mu_x$ and $\mu_y$ relative to the mean $x$ and $y$ of the template LFC.  We then fit for the transformation parameters in $x$ and $y$ such that the measured line centers $l_{n,i,x/y}$ for a given LFC exposure $n$ and given LFC line $i$ best match those of the template LFC exposure with centers we will denote $l_{r,i,x/y}$. Lines are first scaled (Equation~\ref{eq:scale}) and then skewed (Equation~\ref{eq:skew}):

\begin{equation}\label{eq:scale}
    l'_{n,i,x} = m_x \cdot l_{n,i,x},
\end{equation}

\begin{equation}\label{eq:skew}
    l_{r,i,x} = l'_{n,i,x} + sin(\theta_{x}) \cdot l'_{n,i,x}
\end{equation}

Identical calculations are performed for both $x$ and $y$ directions using the same model framework. A global scaling and skew value in both x and y are fit for using all LFC line centers fit for in an observation.  The resulting time series of \echellogram\ shifts, skews, and scalings are shown in Figure~\ref{fig:housekeeping}, grouped by transformation type.  The full time series of binned RVs as well as the associated values for the different instrument tracers described here are included with the data published along with this paper.


\section{Correction}\label{sec:methods}
To correct for instrumental systematics not captured by the standard reduction pipeline, we implement a simple multi-dimensional linear regression using a suite of instrumental tracers. These tracers include both telemetry-based quantities, such as temperature sensor readings, and empirically derived parameters, such as those described in Sections~\ref{ssec:lbs} and~\ref{ssec:echellogram}. In this framework, the tracers serve as the independent variables, or regressors, denoted $x_1, x_2, \dots, x_T$. For $N$ observations, we construct the following classic design matrix:

\begin{equation}
    X =
    \begin{bmatrix}
        x_{1,1} & x_{1,2} & \dots & x_{1,T} \\
        x_{2,1} & x_{2,2} & \dots & x_{2,T} \\
        \vdots  & \vdots  & \ddots & \vdots \\
        x_{N,1} & x_{N,2} & \dots & x_{N,T}
    \end{bmatrix}.
\end{equation}

With the measured RVs as the dependent variable vector $Y$ and the associated uncertainties represented by the diagonal matrix $C$, the best-fit coefficients, we will denote with teh vector $\hat{\beta}$, are obtained as
\begin{equation}
    \hat{\beta} = (X^{T} C^{-1} X)^{-1} (X^{T} C^{-1} Y),
\end{equation}
yielding model predictions, $\hat{y}$, which give the expected change in measured RV due to instrumental systematics as
\begin{equation}
    \hat{y} = \hat{\beta} X.
\end{equation}

Because the various tracers are sampled at different cadences, all values are interpolated to the times of the RV measurements. For \expres, the temperature sensors are sampled more frequently than the science exposures, while the LFC files that provide empirical tracers typically bracket the nightly observations. To associate each science exposure with corresponding tracer values, we first bin the source data and then interpolate between these bins using a cubic spline. Binning has the added benefit of suppressing high-frequency variations not reflected in the measured RVs.

In the implementation described here, temperature sensors are averaged over four-hour bins, whereas empirically derived quantities are binned daily.  For the over-sampled temperature sensors, the bin width was primarily determined by the width at which the Spearman correlation coefficient between the binned temperature time series and the binned Solar RVs plateaued.  That signified that the higher-frequency signals were being appropriately smoothed over.  For the book-end empirical values, we tested bin widths of four hours up to three days.  For both, the final bin width was chosen to minimize the difference between the interpolated value using the binned values and the original, non-binned values.

Before fitting, all tracer values are normalized to zero mean and unit variance to account for their differing units and scales.  We use the binned RVs (median-filtered over 11 days) as the dependent variable to cut down on astrophysical sources of noise that are not shared between all targets.  To guard against overfitting, the model coefficients are derived separately for each target, with that target’s data excluded from the training set. In the case of $\Delta$ solar data, 10\% of the observations are randomly withheld during training.

Figure~\ref{fig:beta} shows the distributions of the resulting best-fit coefficients across all training and validation sets. Even when different targets or subsets of $\Delta$ solar data are withheld, the fitted coefficients remain tightly clustered, demonstrating the robustness of the correction model.  The exact value of the coefficient is expected to differ between the night and $\Delta$ solar fits since they are performed independently, span different time ranges, and the solar residuals will also contain \neid\ systematics.

\begin{figure*}[t!]
\centering
\includegraphics[width=.75\textwidth]{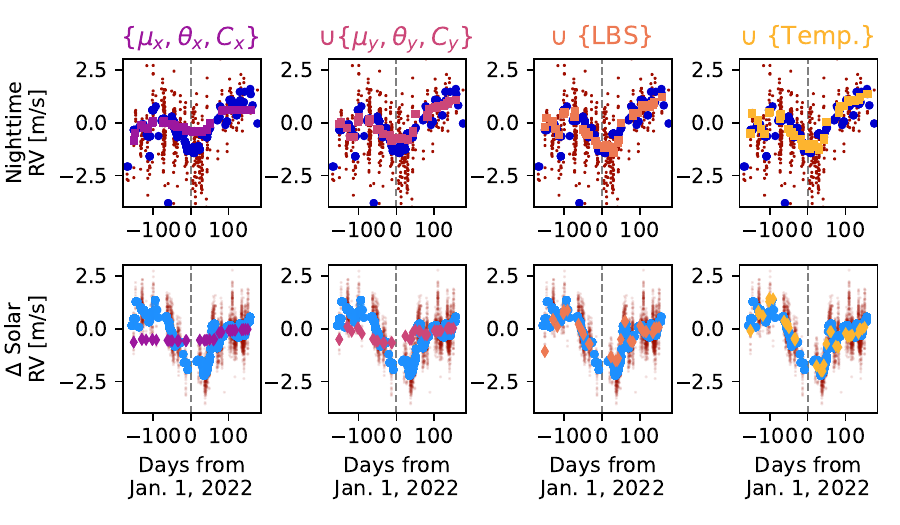}
\caption{Best-fit RV corrections for \expres\ RVs within $\pm$6~months of 2022~January~1 (vertical dashed gray line).  Corrections accumulate left to right, with each column adding the listed tracers and retaining all from prior columns.
\textbf{Top}: Nighttime \expres\ observations (red) binned to emphasize the coherent instrumental trend (dark blue).
\textbf{Bottom}: $\Delta$ Solar observations (red) and binned $\Delta$ solar observations (light blue).
Corrections are over plotted in both rows with square and diamond markers, respectively.}
\label{fig:corrBuildUp}
\end{figure*}

\section{Results}\label{sec:results}

The best-fit RV corrections depend on the set of instrumental tracers included in the model. We assess the efficacy of the corrections using the RMS of the corrected, binned RVs, as binning averages over astrophysical sources of variability. We further evaluate performance through injection/recovery tests of simulated planets and by examining the impact on the previously published fit of $\rho$ Coronae Borealis (\rcb) using \expres\ data \citep{brewer2023}.

\subsection{Incorporating Different Instrumental Tracers}\label{ssec:tracers}

Figure~\ref{fig:corrBuildUp} demonstrates how the RV corrections evolve as additional instrumental traces are cumulative introduced from left to right. As in earlier figures, we zoom in on the dip in the binned \expres\ data near 2022~January, where the instrumental trend is most apparent. Corrections for the nighttime data (top row) and the $\Delta$ solar data (bottom row) are overplotted on the instrumental trend (dark and light blue, respectively). Corrections that perform well will closely track the instrumental trend. Each successive panel, left to right, adds the additional tracers listed at the top of the column---each column is a superset of all columns to the left of that column.

\begin{figure}[t!]
\centering
\includegraphics[width=.45\textwidth]{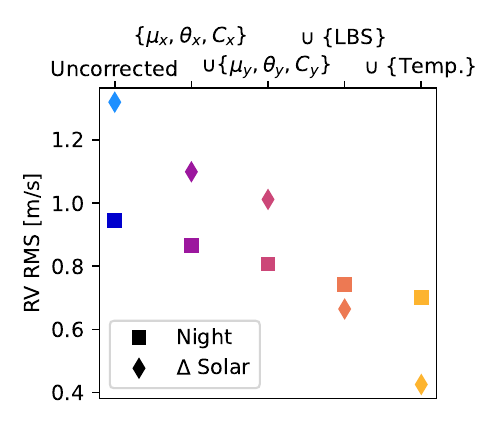}
\caption{RMS of all binned, corrected RVs for models using different sets of instrumental tracers for both nighttime (squares) and $\Delta$ solar data (diamonds).  As with the different rows in Figure~\ref{fig:corrBuildUp}, each subsequent x-axis tick label (left-to-right) is a union of previous and listed tracers---i.e.\ moving left-to-right, increasingly more tracers are incorporated.}
\label{fig:corrRms}
\end{figure}

\begin{figure*}[t]
\centering
\includegraphics[width=.8\textwidth]{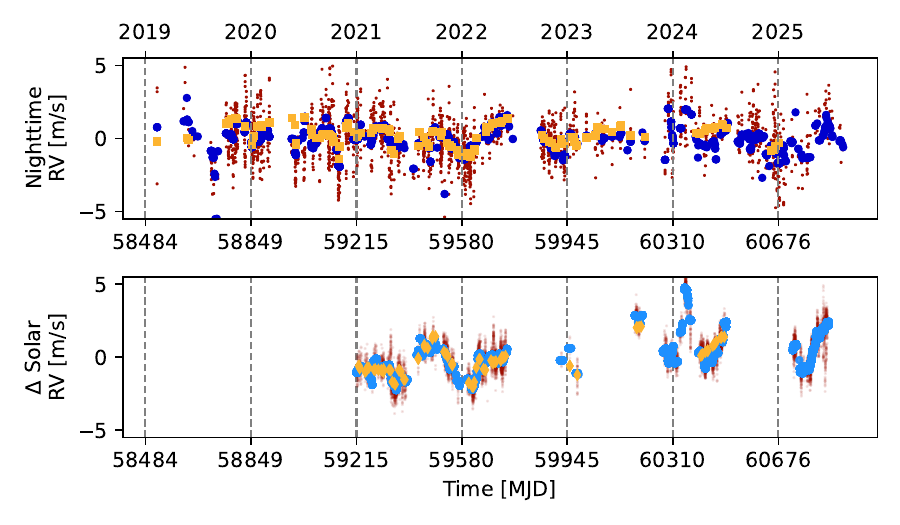}
\caption{Best-fit RV corrections (yellow) using all \echellogram\ position terms ($x$ and $y$), LBS measurements, and temperature telemetry. Vertical dashed gray lines mark the beginning of each calendar year from 2019 to 2025. Corrections are shown over the original RVs (red) and the binned RVs (blue).}
\label{fig:corrections}
\end{figure*}

When the model includes only \echellogram\ position changes in the dispersion ($x$) direction (Figure~\ref{fig:corrBuildUp}, first column), the corrections do not closely follow the instrumental trend, likely because most $x$-direction changes are already captured when fitting calibration line positions. Adding \echellogram\ position changes in the cross-dispersion ($y$) direction (second column) yields corrections that more faithfully trace the observed trend in the nighttime data (top row).  While the $x$ and $y$ direction shifts trace the same large offsets (see Figure~\ref{fig:housekeeping}), incorporating a direct measure of shifts in the cross-dispersion direction is more likely to capture variations missed by the existing wavelength calibration.  Incorporating LBS values (third column) introduces sensitivity to IP asymmetry, which improves the corrections for the solar data. Finally, including temperature telemetry (fourth column), although not a direct spectral tracer, helps account for additional instrumental variations not captured by the other diagnostics for both data sets.  The slightly different behavior between the nighttime data and the solar data is to be expected given the different underlying signals in the two data sets (i.e.\ undetected planets vs.\ \neid\ instrument trends respectively).

Figure~\ref{fig:corrRms} summarizes the RMS of the binned, corrected RVs across all available data for the same tracer subsets as shown in Figure~\ref{fig:corrBuildUp}. Our simulations indicate that the binned \expres\ data should average down planetary signals to $\sigma \lesssim 70~\cms$, so the RMS of the binned, corrected RVs serves as a reasonable proxy for correction performance. Using both $x$ and $y$ \echellogram\ position terms, LBS, and temperature telemetry yields the lowest RMS. Indeed, the resultant RMS of the nighttime data is the expected 70.8~\cms\ while the $\Delta$ solar data, for which Solar System planets have been removed, have a final binned RV RMS of 44.0~\cms.  In the following analyses we therefore adopt corrections from the full suite of tracers. Figure~\ref{fig:corrections} shows these corrections over the full \expres\ time span.

\subsection{Injection/Recovery Analysis}

To verify that the correction procedure does not absorb planetary signals, we perform injection/recovery tests over a grid of planet masses and periods. For these tests, we consider single-planet systems with zero eccentricity. Planet masses range from 1 to 25~$M_{\oplus}$ and orbital periods from 3~days to 1~year; this corresponds to semi-amplitudes of $K \sim 0.01$–$20~\ms$ for a solar-mass host.

For nighttime data, we inject simulated planet RVs into the measured RVs of each of the twelve low-RMS \expres\ targets. This preserves realistic sampling, since we use the actual time stamps of the survey for each target. The resulting RVs therefore contain the injected planetary signal, the \expres\ instrumental trend, and any real stellar or planetary signals present in the target data.

For the solar tests, we sample from the \expres$-$\neid\ solar, i.e.\ $\Delta$ solar, RV time series. These RVs thus contain the injected planet signal, the \expres\ instrumental trend, any \neid\ trends, and possible residual solar signals present in one but not both instruments. To approximate a realistic cadence, we randomly select 300 observations from the available $\Delta$ solar data and generate 12 independent $\Delta$ solar subsets (each with 300 observations), analogous to the twelve nighttime targets.

For each of these 24 time series, we inject a planet with a given mass and period.  We then re-calculate the full correction model using the time series of RVs that now include the injected planet signal.  Each injected planet signal is therefore independently fit to the correction model.  We then compute a Lomb-Scargle periodogram of these corrected RVs.  A planet is considered ``detected'' if the maximum-power period lies within 15\% of the injected period.

\begin{figure*}[t!]
\centering
\includegraphics[width=.8\textwidth]{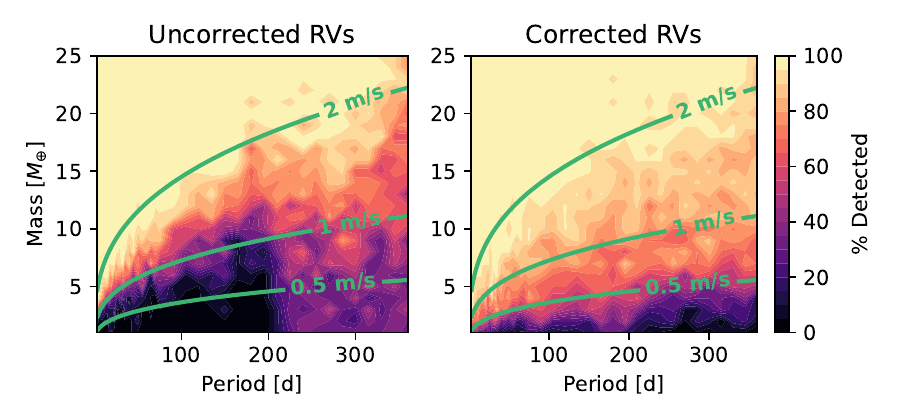}
\includegraphics[width=.8\textwidth]{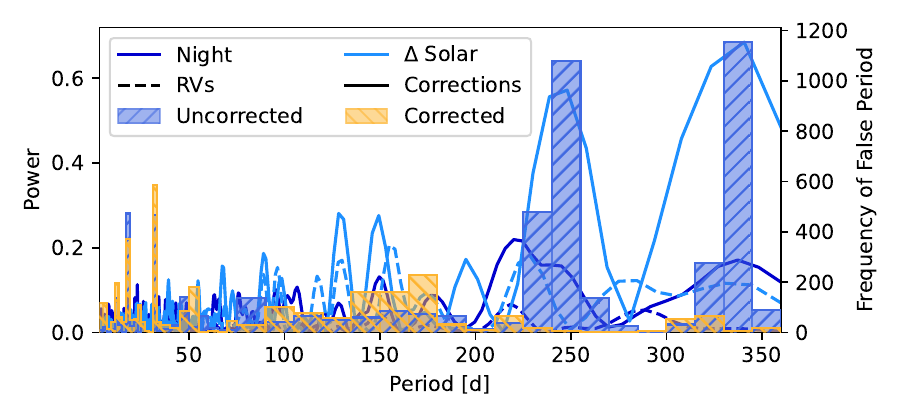}
\caption{\textbf{Top}: Fraction of injected planets recovered as a function of mass and period in uncorrected (left) and corrected (right) data. A detection requires the maximum-power period in the RV periodogram to be within 15\% of the injected period. Green curves show constant semi-amplitude $K$ for a solar-mass host.
\textbf{Bottom}: Lomb–Scargle periodograms of the binned nighttime and $\Delta$ solar RVs (solid lines), together with periodograms of the best-fit corrections (dashed lines). Overplotted histograms show the distributions of false periods, i.e., the maximum-power period when it is not within 15\% of the injected value.}
\label{fig:ir}
\end{figure*}

Figure~\ref{fig:ir} shows the fraction of detections as a function of injected mass and period for the uncorrected (left) and corrected (right) RVs. Lower-mass planets down to $K \sim 0.5~\ms$ with $P \lesssim 200$~days are detected more consistently after correction, indicating that the procedure preserves true Doppler signals and improves sensitivity to low-amplitude planets.

In the uncorrected data, periodograms frequently return maximum-power periods between 200 and 300~days. Because longer-period planets generally produce lower-amplitude signals; the apparent increase in correct-period identifications in this range is unlikely to reflect true planetary detections. Instead, we compare with the periodograms of the binned RVs, which more directly reflect the instrumental trend, and with the periodograms of the best-fit corrections (bottom panel of Figure~\ref{fig:ir}).

The corrections exhibit strong power at $P \gtrsim 200$~days. The overplotted histograms show the distributions of maximum-power periods for cases where the injected period is not recovered. In the uncorrected data (blue), these false periods cluster at the same long periods as the instrumental trend, consistent with spurious detections driven by instrumental periodicities. In the corrected data, the distribution of false periods (yellow) is substantially diminished at these periods, indicating that the correction successfully mitigates the periodic instrumental trend.




\subsection{$\rho$ Coronae Borealis (HD~143761)}

HD~143761, or $\rho$~Coronae Borealis (\rcb), is a G0 star ($V=5.39$~mag). Four planets have been reported around \rcb: a 39-day hot Jupiter detected with AFOE~\citep{noyes1997}; a 102-day, 25~$M_{\oplus}$ planet from Keck/HIRES~\citep{fulton2016}; and two additional planets from \expres\ data at 281.4~days (20~$M_{\oplus}$) and 12.95~days (3.7~$M_{\oplus}$)~\citep{brewer2023}.

\begin{figure*}[t!]
\centering
\includegraphics[width=.9\textwidth]{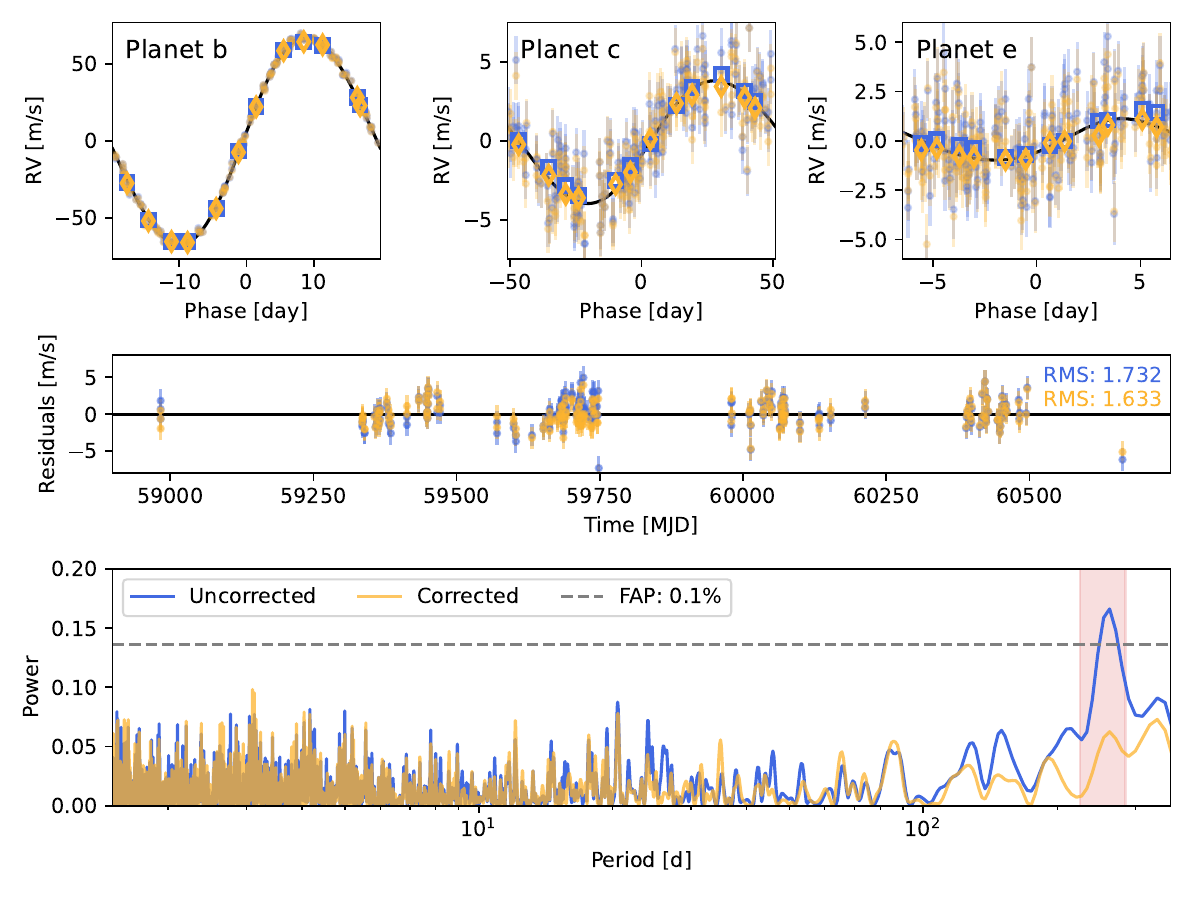}
\caption{\textbf{Top}: Uncorrected (blue) and corrected (yellow) RVs phase-folded to the three \rcb\ planets that remain significant with the corrected RVs.  Squares and diamonds show phase-binned RVs for uncorrected and corrected RVs respectively.  
\textbf{Bottom}: Lomb-Scargle periodograms of the residuals of the uncorrected and corrected data to the three-planet fit.  Red shaded region highlights periods between 225 and 285~days, which showed an increase in false-positive planet detections in injection/recovery tests.}
\label{fig:rcb}
\end{figure*}

Here, we use 253 \expres\ observations of \rcb\ taken on 197 different nights from 2020~May to 2024~December.  Figure~\ref{fig:rcb} shows the uncorrected and corrected RVs phase-folded to the three reported planetary signals that remain significant.  We see that the final residuals for this three-planet fit decreases from 1.732~\ms\ to 1.618~\ms.  The larger impact is in the frequency space, as can be seen in the bottom row of Fig.~\ref{fig:rcb}, which shows the periodograms of the residuals of the uncorrected (blue) and corrected (yellow) RVs to a  three-planet fit.  In the residuals of the uncorrected data, there persists a significant signal with a peak at 262.48~days that is not present in the residuals of the corrected data.   The need for this correction would not have been obvious from the residual RMS alone.

\begin{table}[tb]
\scriptsize
\caption{\rcb\ RVs}
\label{tab:rcb_data}
\begin{center}
\begin{tabular}{c c c c}
\hline
\hline
Time [MJD] & Uncorrected & Corrected & RV Error \\
 & RV [\ms] & RV [\ms] & [\ms] \\
\hline
\input{rcb_bin-tempEmpr_rvTable}
\end{tabular}
\end{center}
\end{table}

\begin{table}[tb]
\scriptsize
\caption{\rcb\ Best-Fit Parameters}
\label{tab:rcb_fit}
\begin{center}
\begin{tabular}{l c c}
\hline
\hline
Parameter & Value & Unit \\
\hline
\input{251015_HD143761fit_bin-tempEmpr}
\vspace{-36pt}
\end{tabular}
\end{center}
\end{table}

Our injection/recovery tests indicate that prior to correcting long-term instrumental systematics, the most common false-positive periods were for injected planets with periods between 225 and 285~days.  This likely explains the previously reported 281.4-day signal in the \expres\ analysis, which is no longer significant after applying the corrections.  The uncorrected and corrected RVs for \rcb\ are given in Table~\ref{tab:rcb_data}.  The best-fit parameters to the corrected RVs, including system-wide RV offset ($RV_0$) and stellar scatter ($\sigma_\star$) terms, are given in Table~\ref{tab:rcb_fit}.

\section{Discussion}\label{sec:discussion}

The 10–30~\cms\ precision goal of extreme-precision radial velocity (EPRV) work demands increasingly careful treatment of instrumental systematics, particularly those that persist beyond standard extraction pipelines. It is no longer sufficient to assume that all relevant instrumental variations can be captured by fitting calibration line positions to a static, symmetric instrumental profile (IP) along only the dispersion direction. As a pertinent example, recent results from \espresso\ demonstrated drastic improvements to their wavelength calibration by better parameterizing their IP.  EPRV measurements require new approaches to identify, quantify, and correct subtle instrumental variations that now limit attainable precision.

We have presented evidence of a coherent instrumental trend in seven years of \expres\ observations. This trend demonstrates that even an ultra-stabilized spectrograph with demonstrated calibration stability of 3–7~\cms~\citep{blackman2020} can exhibit small residual variations that propagate into long-term RV measurements.  The trend only became apparent after accumulating a sufficiently long temporal baseline.  We found that the largest variations coincide with periods of greater temperature variation.  

We describe a set of instrumental tracers that include measures of calibration line asymmetry (LBS \S\ref{ssec:lbs}), two-dimensional \echellogram\ variations (\S\ref{ssec:echellogram}), and the temperature of the instrument.  LBS measurements captured asymmetries in the \expres\ LBS that varied at the scale of $10^{-6}$~nm.  The \echellogram\ can shift in x \emph{and} y up to two pixels.  Much more subtle is the skew of the \echellogram, which is on the order of $10^{-4}$~pixels, and the scale factor, which is typically on the order of 0.999 where 1 is no scaling.  In the EPRV context, even these very small instrumental variations must be taken into account.

These measured variations in the asymmetry of the LFC lines and \echellogram\ positions was unexpected as neither the rectangular fiber face nor the CCD should be susceptible to rotations relative to each other.  The measured changes, however, are small and so are likely to be caused by very subtle variations.  For instance, given that the clearest RV variations presented at a time of greater temperature variation, this could suggest that changes in temperature may have caused very slight tilts or rotations to the camera barrel lenses, causing the observed IP and \echellogram\ changes.  We can only speculate on the exact physical causes of the observed RV shifts, but this does highlight the level of long-term hardware stability needed to carry out EPRV work.

These variations likely translated to RV shifts as they impact the implicit assumptions of the extraction pipeline and they also may affect calibration emission lines differently than the absorption lines in science observations.  Changing IP asymmetry makes it difficult to define a consistent concept of the centroid of a line, which consequently complicates the calculation of the shift of a line.  Classical calibration sources give rise to bright emission lines, which will illicit a different CCD response than the absorption lines of stellar spectra due to considerations like changing CTI and the brighter-fatter effect.

Measuring these variations using different instrument tracers allowed us to implement a simple multi-dimensional linear regression, which substantially improved the integrity of the \expres\ radial velocity data.  In particular, the revised data set no longer supports the existence of the previously reported ``planet~d'' around $\rho$~Coronae Borealis \citep{brewer2023}, providing a clear demonstration that the instrumental corrections preserve genuine Doppler signals while eliminating spurious detections arising from long-term instrumental systematics.  We ran an injection/recovery test of low-amplitude planets to further attest to the robustness of the correction.

The stability and dense, narrow lines of the laser frequency comb (LFC) were key to diagnosing and tracking this systematic. Because individual LFC modes are much narrower than the resolving power of \expres, each can be modeled as a delta function convolved with the IP, allowing them to serve as sensitive probes of instrumental behavior and symmetry changes. Using LFC exposures taken through the science fiber ensures that calibration and stellar light share the same optical path, although the possible influence of signal-dependent charge transfer efficiency (CTE) was not considered in this analysis.

All precision-RV instruments should periodically check for coherent systematics in this manner. RVs from all targets can be binned to test for trends common across the instrument. It is important to explore a range of bin widths (see Figure~\ref{fig:binwid}) and to ensure that each bin includes multiple independent targets (Figure~\ref{fig:binning}) to average over astrophysical noise. Additional diagnostics that are typically absent from classical pipeline corrections--—such as two-dimensional \echellogram\ shifts, skews, and scalings, or changes in IP asymmetry--—can reveal further sources of variation.

Here we report a global value for the different instrument tracers for each observation.  It is possible that variations in the echellogram or IP asymmetry may behave differently across sub-regions of the detector.  Indeed, initial analysis suggests that measured LBS changes slightly across the detector.  Though out of scope for the correction-oriented nature of this work, more detailed investigation of potential changes across the detector is a promising avenue to better characterizing instrumental changes.  For instance, such information could be a valuable addition to hierarchical wavelength calibration models, such as \code{excalibur} \citep{zhao2021}

Implementing the correction described here required careful tuning of both dependent (binned RVs) and independent (instrumental tracers) variables. Selecting an appropriate characteristic timescale, defined here by the bin width, allows smoothing of unrelated short-timescale variability while preserving longer-term instrumental behavior. Omitting important tracers can yield incomplete corrections, while including too many result in a poorly constrained model. Normalizing all tracers to zero mean and unit variance keeps the fit well-conditioned.  Separate training and validation sets are essential to prevent overfitting.

Given the range of possible model configurations, it is important to establish consistent performance metrics. As shown in Section~\ref{ssec:binning}, reductions in the RMS of binned RVs can be used to evaluate correction efficacy up to a limit. The most relevant metric depends on the use case; for planet-search spectrographs, the ultimate test is improvement in the goodness-of-fit to planetary systems.  In an EPRV context, residual RMS may not be a sensitive enough probe.  Injection–recovery tests remain the gold standard for assessing overfitting, though their interpretation can be complicated by undetected planets and stellar activity.

The various instrumental tracers—temperatures, telemetry, and empirically measured optical shifts—can be viewed as the instrumental analogs of stellar ``activity indicators.'' Thus, the more sophisticated methods developed to decorrelate RVs from stellar activity (e.g., Gaussian processes, non-linear regressions, etc.) could also be applied in the instrumental domain. Here we have implemented only the simplest linear model. In our tests, we found no evidence of temporal lags or correlations with time derivatives of the tracers, but these may exist for other systems. Future work could explore phase-invariant models or higher-order terms.

Ultimately, reduction pipelines that intrinsically account for such variations will outperform any post-hoc correction. This may require operating directly on the two-dimensional \echellogram, where positional shifts in both directions are preserved. Accounting for IP asymmetries must be done consistently for both calibration and science lines. This could involve modeling CTE effects or adopting calibration sources that more closely resemble stellar spectra, such as well-characterized telluric or absorption-cell lines.

\section{Conclusion}\label{sec:conclusion}
We have identified and characterized a coherent, long-term instrumental trend in seven years of \expres\ data that was not captured by standard pipeline calibrations. By combining laser frequency comb diagnostics with solar and stellar observations, we have demonstrated that subtle, wavelength- and time-dependent variations in the instrumental profile and \echellogram\ geometry can produce measurable shifts in radial velocity zero points at the meter-per-second level.  

To mitigate these effects, we developed a multi-dimensional linear correction that incorporates both telemetry- and data-derived tracers of the instrument state. Applying this correction reduced the RMS of the \expres\ radial velocity measurements and improved the fidelity of planetary detections, particularly for low-amplitude signals.  Injection–recovery tests confirm that the correction preserves true Doppler signals while suppressing spurious periodicities correlated with thermal variations in the instrument.  The apparent Planet \rcb~d was revealed to be an artifact of these systematics, demonstrating that the corrections enhance the integrity of the \expres\ RV data and strengthen confidence in the long-term stability and reliability of the \expres\ time series.

The results emphasize that even ultra-stable, vacuum-enclosed spectrographs such as \expres\ are subject to residual systematics that can masquerade as astrophysical variability. For next-generation EPRV efforts, long-term monitoring with solar feeds, regular calibration diagnostics, and multi-dimensional modeling of instrumental behavior will be essential. Future pipelines that incorporate these corrections at the two-dimensional \echellogram\ level, accounting for instrumental asymmetries and temporal evolution, will further advance the goal of achieving true centimeter-per-second precision in radial velocity measurements.

\software{SciPy library \citep{scipy}, NumPy \citep{numpy, numpy2}, Astropy \citep{astropy2013,astropy2018}}

\acknowledgements
We thank David W. Hogg (NYU, CCA) and Megan E. Bedell (CCA) for valuable discussions. Support for this work was provided by NASA through the NASA Hubble Fellowship grant HST-HF2-51569 awarded by the Space Telescope Science Institute, which is operated by the Association of Universities for Research in Astronomy, Inc., for NASA, under contract NAS5-26555.

These results made use of the Lowell Discovery Telescope at Lowell Observatory. Lowell is a private, non-profit institution dedicated to astrophysical research and public appreciation of astronomy and operates the \ldt\ in partnership with Boston University, the University of Maryland, the University of Toledo, Northern Arizona University and Yale University.  Lowell Observatory sits at the base of mountains sacred to tribes throughout the region.  We honor their past, present, and future generations who have lived here for millennia and will forever call this place home.

The \expres\ team acknowledges support for the design and construction of \expres\ from NSF MRI-1429365, NSF ATI-1509436 and Yale University.  We are also grateful for two grants from the Mt.\ Cuba Astronomical Foundation to support needed replacement and repairs to \expres\ components, which have enabled its continued operation over the last several years.  DAF gratefully acknowledges support to carry out this research from NSF 2009528, NSF 1616086, NSF AST-2009528, the Heising-Simons Foundation, and an anonymous donor in the Yale alumni community. JMB acknowledges support from NSF 2307467.

This paper contains data taken with the NEID instrument, which was funded by the NASA-NSF Exoplanet Observational Research (NN-EXPLORE) partnership and built by Pennsylvania State University.  We thank the NEID Queue Observers and WIYN Observing Associates for their skillful execution of NEID's nighttime observing programs and careful monitoring of NEID's calibration exposures.

NEID is installed on the WIYN telescope, which is operated by the NSF's National Optical-Infrared Astronomy Research Laboratory (NOIRLab).  The NEID archive is operated by the NASA Exoplanet Science Institute at the California Institute of Technology.  Part of this work was performed for the Jet Propulsion Laboratory, California Institute of Technology, sponsored by the United States Government under the Prime Contract 80NM0018D0004 between Caltech and NASA.

This paper is based in part on observations at Kitt Peak National Observatory, NSF’s NOIRLab, managed by the Association of Universities for Research in Astronomy (AURA) under a cooperative agreement with the National Science Foundation. The authors are honored to be permitted to conduct astronomical research on Iolkam Du’ag (Kitt Peak), a mountain with particular significance to the Tohono O’odham.

Deepest gratitude to Zade Arnold, Joe Davis, Michelle Edwards, John Ehret, Tina Juan, Brian Pisarek, Aaron Rowe, Fred Wortman, the Eastern Area Incident Management Team, and all of the firefighters and air support crew who fought the recent Contreras fire. Against great odds, you saved Kitt Peak National Observatory.

\bibliography{paper}

\end{document}

%% file: customCommands.tex
\newcommand{\project}[1]{\textsl{#1}}
\newcommand{\acronym}[1]{{\small{#1}}}
\newcommand{\code}[1]{\texttt{#1}}

\newcommand{\expres}{\acronym{EXPRES}}

\newcommand{\ldt}{\acronym{LDT}}

\newcommand{\espresso}{\acronym{ESPRESSO}}
\newcommand{\neid}{\acronym{NEID}}
\newcommand{\wiyn}{\acronym{WIYN}}

\newcommand{\cms}{\mbox{cm~s\textsuperscript{-1}}}
\newcommand{\ms}{\mbox{m~s\textsuperscript{-1}}}


%% file: rcb_bin-tempEmpr_rvTable.tex
58983.2365 & 19.9747 & 21.2194 & 0.3996 \\ 
58983.2386 & 21.0853 & 22.3305 & 0.4157 \\ 
58983.2408 & 18.6072 & 19.853 & 0.4261 \\ 
59335.4115 & 52.0664 & 51.4311 & 0.3557 \\ 
59335.4146 & 51.5921 & 50.9576 & 0.3487 \\ 
$\vdots$ & $\vdots$ & $\vdots$ & $\vdots$ \\ 
\hline

%% file: 251015_HD143761fit_bin-tempEmpr.tex
\textbf{Planet b} &  &  \\ 
$P_b$ & 39.849$\pm$0.002 & days \\ 
$T_{peri_b}$ & 55499.084$\pm$0.182 & MJD \\ 
$e_b$ & 0.046$\pm$0.002 &  \\ 
$\omega_b$ & 4.715$\pm$0.05 & rad \\ 
$K_b$ & 67.488$\pm$0.174 & \ms \\ 
$M_b sin i$ & 347.995$\pm$0.054 & $M_{\oplus}$ \\ 
\hline 
\textbf{Planet c} &  &  \\ 
$P_c$ & 102.036$\pm$0.204 & days \\ 
$T_{peri_c}$ & 55480.216$\pm$9.336 & MJD \\ 
$e_c$ & 0.048$\pm$0.034 &  \\ 
$\omega_c$ & 4.265$\pm$0.957 & rad \\ 
$K_c$ & 3.891$\pm$0.16 & \ms \\ 
$M_c sin i$ & 27.442$\pm$0.709 & $M_{\oplus}$ \\ 
\hline 
\textbf{Planet e} &  &  \\ 
$P_e$ & 12.904$\pm$0.011 & days \\ 
$T_{peri_e}$ & 55496.721$\pm$3.641 & MJD \\ 
$e_e$ & 0.082$\pm$0.047 &  \\ 
$\omega_e$ & 5.773$\pm$1.797 & rad \\ 
$K_e$ & 1.05$\pm$0.16 & \ms \\ 
$M_e sin i$ & 3.71$\pm$0.598 & $M_{\oplus}$ \\ 
\hline 
\textbf{System-wide} &  &  \\ 
$RV_0$ & -8.655$\pm$0.109 & \ms \\ 
$\sigma_\star$ & 1.407$\pm$0.0728 & \ms \\ 
\hline